\documentclass[conference]{IEEEtran}
\IEEEoverridecommandlockouts

\usepackage[utf8]{inputenc}
\usepackage[T1]{fontenc}
\usepackage{graphicx}
\usepackage{graphics}
\usepackage{xcolor}
\usepackage[normalem]{ulem} 
\usepackage[official]{eurosym}
\usepackage[switch,modulo]{lineno}
\usepackage[normalem]{ulem} 
\usepackage{footnote}

\usepackage{amsmath}

\usepackage{hyperref}  

\makeatletter
\newcommand{\linebreakand}{%
      \end{@IEEEauthorhalign}
      \hfill\mbox{}\par
      \mbox{}\hfill\begin{@IEEEauthorhalign}
    }
    \makeatother

\title{\LARGE \textbf{Efficiency in Digital Economies}\\ -- A Primer on Tokenomics --}

\begin{document}



\author{%
	\IEEEauthorblockN{\textbf{Ricky Lamberty}}
	\IEEEauthorblockA{%
		\textit{University of Digital Science}\\
		\textit{Robert Bosch GmbH}\\
		Potsdam \& Stuttgart, Germany\\
		ricky.lamberty@de.bosch.com}
	    \and
	\IEEEauthorblockN{\textbf{Alexander Poddey}}
	\IEEEauthorblockA{%
		\textit{Robert Bosch GmbH}\\
		Renningen, Germany\\
		alexander.poddey@de.bosch.com}
		\and
	\IEEEauthorblockN{\textbf{David Galindo}}
	\IEEEauthorblockA{%
		\textit{Valory AG}\\
		Zug, Switzerland\\
		david.galindo@valory.xyz}
    \linebreakand
    \IEEEauthorblockN{\textbf{Danny de Waard}}
    \IEEEauthorblockA{%
    	\textit{Robert Bosch GmbH}\\
    	Renningen, Germany\\
    	danny.de.waard@de.bosch.com}
    \and
  	\IEEEauthorblockN{\textbf{Tobias Kölbel}}
    \IEEEauthorblockA{%
    	\textit{Karlsruhe Institute of Technology}\\
    	\textit{Robert Bosch GmbH}\\
    	Karlsruhe \& Renningen, Germany\\
    	tobias.koelbel@de.bosch.com}
    \and
  	\IEEEauthorblockN{\textbf{Daniel Kirste}}
    \IEEEauthorblockA{%
    	\textit{Robert Bosch GmbH}\\
        Stuttgart, Germany\\
    	daniel.kirste@de.bosch.com}}

\maketitle
\thispagestyle{plain}
\pagestyle{plain}


    \begin{abstract}


Cryptographic tokens are a new digital paradigm that can facilitate the establishment of economic incentives in digital ecoystems. Tokens can be leveraged for the coordination, optimization and governance of large networks at scale in a decentralized manner. A key aspect is their programmability, that can reward participants relative to their stage of adoption, according to the value they contribute and the risk they bear. 
Moreover, this can be done in a transparent and verifiable way, which increases trustworthiness in the emerging systems. This work presents an overview of this new phenomenon and to provide multi-disciplinary arguments on why tokenized ecosystems can drive a huge momentum for positive-sum collaboration in the digital age. We illustrate how certain principles and values that arise from the evolutionary process of digital cooperation can lead to a market economy characterized by economic efficiency of both individuals \textit{and} the tokenized ecosystem as a whole.

\end{abstract}
    
\section{Introduction}


\emph{Distributed Ledger Technology (DLT)} provides an organizationally decentralized approach for maintaining shared data states that bear the potential to transform the "\emph{Internet of Information}" into the "\emph{Internet of Value}" \cite{tasca2020}. The {Internet of Information} has become a major foundation of today's socio-economic systems by enabling the creation and exchange of information at a scale previously unknown. 
Cryptographically secure tokens represent a digital-native approach to credible scarcity and thereby hold value within digital networks such as Bitcoin \cite{galloway2021}. 


Organizationally decentralized but logically centralized\footnote{Namely, the same logic applies in every part of an otherwise decentralised system.}, the maintenance of shared tamper resistant and publicly accessible states allow the creation of software protocols that can undermine the power of \emph{centralized intermediaries}. 

Bitcoin was the first cryptocurrency to introduce a robust and distributed consensus mechanism, known as "\emph{Proof-of-Work}" (PoW). This mechanism was designed in a way that it made the economic cost of attacking the system disproportionate to the benefit of doing so. Bitcoin's PoW consensus mechanism has been proven over ten years of use to be an effective tool against censorship and counterfeiting, making these actions economically unprofitable \cite{granot2019}. The \emph{tokenized} incentive mechanism of the Bitcoin blockchain has not only secured the network but has also coordinated its participants, making it a practical and successful application of DLT.

Cryptographically and game-theoretically secured distributed consensus has sparked a new field of science around economic coordination, sometimes referred to as “\emph{Cryptoeconomics}” or "\emph{Tokenomics}". It can be described as the study of efficient economic interaction and coordination in untrusted environments that are beneficial to the participants of the network. Although, in principle every participant could be corrupt, it can be shown that sensible economic interaction is still possible \cite{shermin2019}. 

In this work we would like to shed light into these novel concepts for a non-cryptographic audience and to provide a clear understanding of the concept and power of \emph{Tokenomics} and its importance for the emerging \emph{digital socio-economy}, while providing only necessary technical details of DLT.


    \section{Fundamentals}

\subsection{Distributed ledger technology}

Distributed Ledger Technology is the technology that enables the realization of distributed ledgers. Distributed ledgers are distributed databases and \emph{blockchain} is a certain form of a distributed ledger. A distributed ledger comprises of a distributed database, but varies from the classic distributed database architecture in three different ways\cite{crosby2016}:

\begin{enumerate}
	\item \emph{Decentralization}: Control is decentralized through multiple or all network participants (peer-to-peer) and has no central point-of-failure.
	\item \emph{Robustness}: The consensus mechanism ensures the integrity of the database, even if the participants do not entirely trust each other.
	\item \emph{Cryptography} is used to ensure (1) and (2) above \cite{benos2017}.
\end{enumerate}

According to Garay \emph{et al.} a decentralized transaction ledger aims at keeping a record of (non-) monetary accounts and its associated balance where a transaction record in the ledger is typically (but not limited to) an instruction to move balances between accounts \cite{garay2015}. Blockchains are one of the most well known concepts of DLT, in which the ledger comprises "blocks" of transactions. The technology functions via a peer-to-peer network, represented by thousands of nodes, e.g. computers, worldwide that run the underlying consensus protocol. DLT enables the creation of cryptographic assets (e.g. Cryptocurrencies), \emph{smart contracts} and the raise of \emph{decentralized (autonomous) organizations} (DAOs) that provide novel \emph{collaboration frameworks} \cite{wright2015}.

\subsection{Game-theory}

Game theory is a study of mathematical models and tools of strategic interaction among rational decision-makers. It is the formal study of conflict and cooperation, where the interests of players may conflict at some situation and cooperate in others. Game-Theory was used to tackle knowledge transfer among rival organizations \cite{samieh2007}. 

In a game, each decision-maker as a player chooses its strategy to maximize its utility, given the other players strategic choice. Thus, game theory can be used to analyze the strategies of blockchain nodes as well as the interactions among them. Through game theoretical analysis, the nodes can predict mining behaviors of each other, then forming optimal reaction strategies based on equilibrium analysis. Moreover, game theory is utilized to develop incentive mechanisms that discourage the nodes from executing misbehavior or launching attacks. As such, game theory is inevitable in e.g. the decision making of the consensus nodes (e.g. miners) running a DLT-based protocol \cite{han2012}. 



The following sub-sections touch the most present game-theoretical fundamentals required for designing and building the incentive-mechanisms underlying and steering \emph{tokenized} protocols/networks.

\subsubsection{Non-cooperative game}

A \emph{non-cooperative game} studies and models the interaction and conflicts among economic agents, where the payoffs of each economic agent depend not only on its own actions, but also on the behavior of other agents. Solution and coordination proposals arising from the field of \emph{game theory} provides a useful understanding among competition between economic agents under strategic interpendency.

In a non-cooperative game, the players do not cooperate by forming coalitions or by reaching agreements. In general, the term non-cooperative means that any cooperation which might arise must be with no communication of strategies among the players. In other words, the strategy that the player takes must be spontaneous, and each player is rational \cite{khan2002}. Examine a public blockchain network such as the \emph{Bitcoin} blockchain. Players, known as miners, strategically buy and invest in computational power to compete for the incentive reward from mining successfully blocks. The miners are rational and the non-cooperative game can be used to model the interaction among those miners \cite{Survey2019}.

\subsubsection{Extensive-form game}

The rules of an extensive-form game are described in such a way, that the agents execute their moves consecutively. 

The aforementioned non-cooperative game can be used to analyze a static game, i.e., the game that has no notion of time and no player has any knowledge of other players actions in advance, and the dynamic game, i.e., the game in which the players strategies are made following a certain predefined order. The dynamic game can be represented in an extensive form to illustrate the sequence of players possible moves, their choices at every decision point, information that each player has about the other players moves, and their payoffs for all possible game outcomes \cite{kroer2014}. 

Considering the scenario of a fork chain selection, the miner chooses a certain chain to mine on at the beginning of every round of mining competition, given the actions taken by the other players in previous mining rounds. At some points, the blockchain forks and leads to the structure similar to a branching tree. Thus, the extensive-form game can be applied for the analysis on which of two chains the players gonna mine \cite{Survey2019}.


\subsubsection{Cooperative Game}

Cooperative game theory assumes that groups of players, called \emph{coalitions}, are the primary units of decision-making, and may enforce cooperative behavior. Consequently, cooperative games can be seen as a competition between coalitions of players, rather than between individual players. The basic assumption in cooperative game theory is that the grand coalition, the group consisting of all players, will form \cite{shapley1953}. One of the main research questions in cooperative game theory is how to allocate in a fair way the payoff of the grand coalition among the players. The answer to this question is related to a solution concept which, roughly speaking, is a vector that represents the allocation to each player. Different solution concepts based on different notions of fairness have been proposed in the cooperative game theory literature \cite{chandrasekaran1994}. One of the most known solution concepts is the \emph{Shapley value}, which has been mathematically solved more than half a century ago by Lloyd Shapley (1953) \cite{shapley1953}. 

Ideally, cooperative games provide a positive-sum game to create a win-win situation for the players. This can be used to model the tension between selfish incentives and community-based shared benefits. We illustrate more on the concepts of Shapley (1953)\cite{shapley1953} and how it effects socio-economic efficiency more in detail in section \ref{coalition}.

\subsection{Terminology}

We start off by introducing some terminology, which will be examined more in detail in subsequent chapters. We refer to the participants of a \emph{cryptoeconomic system} as economic agents. Economic agents might either be human or artificial individuals trying to connect in order to produce, create, exchange and communicate within a market.

\subsubsection{Smart contract}

A \emph{smart contract} is code that exists and is executed on the distributed ledger when predetermined conditions are met. They are like computer programs that consist of a set of rules and are distributed across the ledger. Smart contracts are used to automate the execution of an agreement such as to receive, store and redistribute tokens without any intermediary's involvement \cite{mohanta2018}.

\subsubsection{Cryptographic token}

Generally speaking, a token is a piece of data which serves as a visible, tangible or intangible representation of an information or a right. For example, a driving license card is a token that represents the fact that you are trained and allowed to drive a car \cite{oliveira2018}. A \emph{cryptographic token}, or \emph{cryptographic asset} in general, is a digital, cryptographically secure, provable representation of an asset, a fact or right, which can be processed in distributed ledger protocols (e.g. a blockchain) \cite{nakamoto2008}. Tokens are digitized multi-purpose instruments, ranging from simple to complex design patterns. Those token could represent value, stake or voting right for instance. A token is not limited to one specific role or utility, it can fulfill a lot of roles in its underlying ecosystem. The process of digital representation of a tangible or intangible asset via a cryptographic token, is called \emph{tokenization} \cite{oecd2020}, which should be seen as a tool or business construct \cite{micro2019}.

\subsubsection{Cryptoeconomics}


Cryptoeconomics can be understood as the combination of cryptography, game-theory and mechanism design to build robust decentralized peer-to-peer networks \cite{zargham2020}. Cryptoeconomic systems, e.g. DLT, provide a public infrastructure that allow the issuance and management of cryptographic tokens and state balances. Cryptography is used to secure the network, game-theory is used to design the interaction and to analyze strategic behavior of economic agents that are interlinked with financial incentives to encourage desired properties. The code of the underlying network is therefore intrinsically interlinked with the economics and incentives of the network \cite{horne2018}. The underlying challenge is that in decentralized P2P systems, that do not give control to any centralized party, one must assume that there will be bad actors lurking to disrupt the system. 



One could consider, that cryptoeconomics mainly focuses on the monetary-aspects of the system, whereas tokenomics has broader aspects than remunerative incentives, like voting rights or network externalities. Due to this, we use from now on the term \emph{Tokenomics} and consider the aspects of \emph{Cryptoeconomics} within this term, which will be furthermore explained in Section \ref{tokenomics}.

\subsubsection{Collaborative data space}

\emph{Collaborative Data Space} (CDS) refers to a relationship between trusted partners who adhere to the same standards and guidelines in relation to data storage and sharing within an ecosystem. Those initiatives are usually shaped by publicly funded consortia contributing open source code\cite{ids2018}. In CDS, partners jointly build the orchestration infrastructure and compete on the application layer \cite{Otto2019}. The main goals are to offer low-cost and large access solutions to counterfeit the power of hyperscalers (such as Google or Facebook)\cite{chesbrough2021}. GAIA-X might become a prime example for a cooperatively operated digital platform in Europe\cite{gaia2022}. It is to be collectively owned across companies and not controlled by a monopolistic provider. The platform is to be developed, operated, and orchestrated cooperatively. 

After laying out the fundamentals and terminology, the next section will look on the main forces, arising from the fields of game-theory, inherently integrated into public and permissionless DLT protocols.

\section{Socio-economic transformation in the digital age}


DLT and related innovations are part of a more general trend towards digitization, computerization, and automation. In a world with growing numbers of important processes rely on formal rules of protocols, the trustworthiness and alignment to more efficient coordination in the context of growing complexity is inevitable \cite{christakis2021}. 

Collaboration frameworks based on DLT and fueled by \emph{tokenized} incentives foster cooperation in the digital age. DLT enables networks in a distributed and decentralized manner, thereby minimizing principal-agent dilemmas \cite{stiglitz1989} of organizations and subsequent moral hazards \cite{marhsall1976} by introducing digital incentive mechanisms. These networks can be compared with digital representations of society and economy, why it is called \emph{socio-economy}. Cryptographic token, within those distributed networks, epitomize those digital incentives to automatically align and enforce interests in the absence of intermediaries. 

Such digital networks are cooperatively owned and governed by their users, as illustrated with CDS. Advocates of digital cooperatives say they are more resilient, more equitable and more sustainable than their centralized, monolithic counterparts. 


\subsection{Web3 and the economy of things}
\label{eot}

The term \emph{Economy of Things} (EoT) evolved from \emph{Internet of Things} (IoT) \cite{ibm2015}. IoT refers to the fact that nowadays, due to ubiquitous connectivity, not only humans connect via the web. It is also possible to build networks of things like sensors, network nodes, cars - so called IoT devices. However, connecting everything with everything by itself is not sufficient. The connected entities need to be able to interact in ways comparable to established economic mechanisms such as search and find, negotiation, payment, settlement, building trust etc. in order to make use of the connectivity \cite{poddey2019}.

The IoT needs to be converted into an EoT. However, although broadly used, the trailing part "of things" is misleading. In fact, what is meant by EoT is a \emph{digital Economy of Everything} (dEoE) - a heterogeneous mix of e.g. small IoT devices, more powerful digital entities like machine learning based services running in the cloud and humans, interacting with each other seamlessly \cite{poddey2019}. 

In Web2, functionality is mostly centralized. Web2 established powerful intermediaries such as Google, Amazon, Facebook, Apple. The functions of a service could in fact be split into several modules, but there typically is a central point of service providing access \cite{gehl2012}. In Web3 even smaller modules can be incorporated as individual entities, connecting and interacting with others on their own behalf. These entities are usually called \emph{agents} \cite{floros2019}. 

Web3 therefore can be understood as a multi-agent system in which functionality - or in a more general form the capability to achieve a goal - emerges from interaction of fragmentary contributions. The capability therefore is no longer embodied in a monolithic entity, but distributed across a network of software agents, each embedding only a part of the necessary modules. Agents providing new fragmentary contributions might appear, others disappear \cite{minarsch2020}. 

This will result in an open digital economy consisting of multiple, co-existing networks. Therefore, the \emph{dEoE} is a prime example of a complex open context system where DLT can provide the soil for digital, economic interaction.

\subsection{Protocols as efficient exchange coordinators}
\label{sec:protocols}

Traditional digital business models are usually built on centralization and the power of a single hub. Centralized businesses are designed to create sales and user numbers to extract as much value as possible. As users increase, profit margins increase, and the business grows around the platform. Users remain disconnected from each other, but inextricably connected to the platform to generate more profit. Due to this, centralized platform raise the issue of personal privacy, ownership of information, efficiency and data protection \cite{christakis2021}\cite{fox2019}.

Generally, protocols can be described as a set of rules or procedures that govern the transfer of data between two or more electronic devices. The protocol defines how data is structured, or how the data is send from one to another party. Protocols allow computer and machines to understand each other, comparable to agreements about language. Usually, those data is used to described certain states between its participants and those states reflect the common knowledge of all involved parties \cite{protocol2020}. 

The power of crypto-based, self-enforcing protocols, such as the Bitcoin protocol, create a new kind of business logic that is revolutionizing a number of industries. While those DLT-based protocols are still in their infancies, it's important to understand why some emerging protocols offer better solutions to the coordination problem. 

Protocols encode the rules of engagement and thereby facilitate fluid economic interaction as systems of logic that coordinate exchange between suppliers and consumers. DLT-based protocols enable connectivity between users in a way that makes sharing of information and value efficient, seamless and even features privacy \cite{santeri2018}. Participants must strictly abide by the rules of engagement, otherwise there will be no exchange. Therefore, a protocol disables or at least reduces human corruption. The parity with which a protocol treats every actor that is connected to it, is part of what drives its efficiency as an exchange coordinator. As exchange coordinator, a protocol should be minimally extractive for its "user", whereas businesses are incentivized, through e.g. shareholders, to be maximally extractive. A protocol may generate a small income for those people operating and mantaining the code, but still re-distributing profits to users throughout the network rather than to a centralized entity. Protocols are in between supplier and consumer, but are no classical centralized  intermediary, that's why they are less extractive. In the absence of a central party, as in the case with DLT-based protocols, protocols provide structures for businesses, but are no businesses per se \cite{burniske2019}.

In direct comparison of protocols with companies facilitating exchange, key differences can be recognized as illustrated in the following table. 



\begin{table}[h!]
\centering
	\begin{tabular}{|l|l|}
		\hline
		\textbf{Company coordinated exchange} & \textbf{Protocol coordinated exchange} \\ \hline
		Consumer pays more                    & Consumer pays less                     \\ \hline
		Company chosen distributor            & Market chosen distributor              \\ \hline
		Company chosen supplier               & Market chosen supplier                 \\ \hline
		\textbf{Maximally extractive}         & \textbf{Minimally extractive}          \\ \hline
	\end{tabular} \\
 
	\caption{Key differences between company- or protocol-coordinated exchange according to Burniske (2019) \cite{burniske2019}.}
	\label{fig:types}
\end{table}

Any supplier can plug into a permissionless and public protocol, so too can any distributor. Therefore, both suppliers and distributors are subject to market competition, as opposed to the proprietary selection process that a centralized company goes through for its suppliers and distributors. Competitive markets kill inefficiencies and drive down costs, which should allow protocol-coordinated services to outcompete company-coordinated services, accruing to the consumer’s benefit \cite{kuznetsov2017}.

Protocols enable and create networks of exchange, as the TCP/IP created the Internet. Instead of being freely open limitlessness, like the Internet, those DLT-based protocols now have \emph{limits}. Limits are introduced through scarce token that store the value of the whole protocol. Those limited protocols now create, enable and entail economic rules such as \emph{tokenized} incentives to maintain its goal \cite{burniske2019}.

\subsection{The delta to traditional economies}

Traditionally, large parts of our society have been organized and secured by a legal system enforcing contractual agreements which regulates the interactions within societies \cite{zumbansen2007}. The governance rules hereby regulate the process of decision making among all stakeholders and allows followers to interact within a community, network or an organization on a sound basis. Rules enable economic exchange and social exchange. Those rules are enforced by the government and its agencies, which provide a central-orchestrated authority \cite{blankart2000}. 

Within a permissionless DLT-based protocol, no central authority exists to enforce those rules. The protocol itself automatically enforces the governance rules set by the community. As such, DLT-based protocols are comparable to the constitution and laws of nation states. Different researchers in economic science, e.g. Stiglitz (1989) \cite{stiglitz1989}, have engaged to the proposition, that markets do not work in a simple and idealized way and therefore nation states mainly steer the actions of their citizens by disincentivize. If you break the law, you either pay or go to jail\cite{atzori2017}. Similar mechanisms, e.g. burning the staked funds\footnote{Staking simply stands for holding a cryptocurrency in a wallet for a fixed period, then earning interest on it.}, are hard-coded into a blockchain protocol as with "Proof-of-Stake" \cite{pos2017}. 

Incentivisation uniquely enabled through cryptographic token and self-enforcing rules via \emph{smart contracts} within the network emphasize a radical change, in-depth Section \ref{sec:economics} \cite{buterin2014}. Those networks have the potential to align stakeholders by the consensus rules. A token incentivises individual behavior in a global distributed network to collectively contribute to a common goal. 

To summarize, a DLT-based network can be compared to a digital nation state, where each digital state has their own interest area, rules and desired system goals. The biggest advantages thereby, ensuring trust by \emph{trustless} technologies and automatization through self-enforcing rules and incentives. \emph{Trustless} in the sense that participants do not need to know or trust each other or a central third party to administer informational or monetary exchange. Instead, users are asked to trust the underlying code that make decentralized protocols secure, reliable, and open to use \cite{Harz2018}.

Due to this, we believe that those kind of protocols provide the substrate for the next evolution of the Internet, Web3, as it enables a new form of decentralized human collaboration and offers the building blocks of digital, scalable and resilient collaboration frameworks. As communities start now to shift from the monolithic platforms of Web2 to open, decentralized protocols of Web3, it raises the question: What is the optimal balance between cooperation and competition and how do we apply tokenized incentives in an efficient manner? 

    \section{Coalition games, coopetition and efficiency}
\label{coalition}





In coalition games, a set of competitive players may cooperate (on some aspects) to form a coalition. This might e.g. relate to resource sharing, sharing of development costs for basic technologies or sharing of modular capabilities to achieve more complex capabilities, unlocking a surplus. Put simply, even though there might be some aspects for which the players are in competition and on which the cooperation might have a detrimental effect, in supermodulary system\footnote{\emph{Supermodularity} is related to convex coalition games, and the solution concept introduced by L. Shapley \cite{shapley1953}.}, cooperation pays of and everybody taking part benefits \cite{song2008}. The combination of competition and cooperation is referred to as \emph{coopetition}, which relates to the supermodularity of the underlying system where agents compete in some modules but cooperate in some others \cite{nalebuff2021}. 

Coopetition in supermodular systems generates a surplus in most cases. The larger the set of agents taking part, the larger the total surplus. Therefore, the goal is to achieve what is called the '\emph{grand coalition}'. The \emph{dEoE} tend to be supermodular. Therefore, it is important to establish a coopetition based dEoE that is based on adequate surplus sharing, both, from a viewpoint of most efficient capability increase, as well as socioeconomic sense \cite{poddey2019}.

Based on the fundamentally important work of Polanyi (first published in 1944), whose theories have been applied to modern economic markets and underpinned by Nobel laureates Stiglitz \cite{stiglitz2002}, Akerloff \cite{akerlof1970} and Spence \cite{smith1759}, there is broad acceptance among economic and social theorists that unregulated systems/markets are generally bound to fail. In other words, systems without rules or interventions do not magically lead to efficiency, so that all possible problems do not seem to resolve by themselves, as some proponents postulate (referring to an invisible hand) \cite{altman2006}. On the contrary: the desired efficiency can only be achieved by framing rules based on fundamental values \cite{polanyi2001}.  

At the same time, however, it is precisely the approach taken by the crypto-movement that offers effective mechanisms and measures to incentive the right behavior, to counteract the danger of concentration of power or suppression inherent in digital technologies, if they are applied in a value-based manner. In this way, there is an opportunity to mold a digital socio-economy based on adequate values in such a way that, thanks to the efficiencies achieved, the threat to these values and the detrimental penetration of destructive actors can be counteracted\cite{poddey2019}.

Digital ecosystems are not about naive cooperation, but coopetition. It is important for a functioning market economy to have a healthy competition, e.g. about providing key expertise. Diminishing competition and arising monopolies have a detrimental effect to the total system. Over and above, for complex systems operating in open contexts, diversity i.a. in the form of competence is important for antifragility and hence persistence of the system. Therefore, a competition based diversity on the one hand, balanced by an adequate protection of minorities form the breeding ground for futures contributors of key expertise and hence prevailing efficiency \cite{poddey2019}.

Maximizing the benefit from a local perspective associated to the selfishness of agents is not evil per se, but the reflection of a necessary contribution to efficiency of the total system. It needs to be counterbalanced, such that entity-local maximization leads to maximization of globally desirable outcomes (i.e. socio-economic optimal results)\footnote{For a detailled discussion, we refer to \cite{poddey2019}.}.

To sum up, coopetition fosters a functioning market economy leading to higher efficiency for its participants. Because of this, we need to ensure the right incentives foster collaboration and competition in the digital age. 

    
\section{Tokenomics}
\label{tokenomics}

\emph{Tokenomics} is the science behind \emph{tokenized} incentives and encompasses the concept of economic system and optimization design to incentivize specific behaviors in a community, using tokens to create a self-sustaining ad hoc economy \cite{shermin2019}. Tokenomics applies game theoretic mechanism design in combination with cryptography to coordinate and create robust, decentralized P2P protocols. Mechanism design is a field in economics and game theory that takes an engineering approach in order to design economic mechanisms such as incentives, towards the desired objectives, in environments where players act rationally\cite{zargham2020}. The main characteristics are the following:

\begin{itemize}
	\item Building systems that have certain desired properties
	\item Applying game-theory and economic incentives to encourage the system to hold desired properties in the future
	\item Using cryptography to prove properties about the past (tamper-proof)
\end{itemize}

By introducing scarcity through tokens, protocols allow for (social) coordination in evolving and complex open system formed by a large number of participants towards the desired goal \cite{zargham2020}. Tokenized incentives thereby protect the underlying ledger from tampering through the means of economic balance. This coordination, based on game-theoretical and economical principles embedded in the protocol, evolves the system as a whole towards the desired properties \cite{shermin2019}. 

The cryptography underlying these systems is what makes the networks secure, and the game-theory is what incentivizes all actors to contribute to the purpose of the network so that it continues to evolve over time. The incentive mechanism should be designed in a way, to make the network fault-tolerant and attack-resistant. Beyond, mechanism design leads the system to evolve to the desired properties over time. This allows entities who do not know one another to reliably reach consensus about the right state \cite{buterin2017}. 

One important difference between economics and tokenomics is that economics most often starts with predictive goals, and tokenomics mainly starts with design goals. This provides the possibility to design a ecosystem according to the desired properties and core values in order to avoid e.g. power concentration or plutarchy \cite{tinsman2018}.

\subsection{Cryptography to secure the present and past}

Cryptography aims to create resilient information systems and is a subfield of cryptology, that almost exclusively refers to encryption. Encryption is the process of converting a piece of information (plaintext) into unintelligible text (ciphertext). Cryptography is used to trustfully identify all network actors, that allows transparency while maintaing at the same time privacy to those network actors. It is an important tool for e.g. managing token through wallets, and is an integral part of consensus mechanisms \cite{shermin2019}. 

A consensus mechanism is a set of self-enforcing rules and processes that define how different nodes can reach an agreement on the true state of the network. Proof-of-Work (PoW) is the first implementation of a distributed consensus protocol and is based on expensive computer computation involving hashing (SHA-256), Merkle Tree and P2P networking for creating, broadcasting and verifying blocks on the network \cite{nakamoto2008}. Currently, more efficient consensus protocols such as Proof-of-Stake (PoS) are getting explored and further evolved \cite{buterin2017}. 

In such a setup, if agents play by to the rules, they get rewarded. It is uneconomical to misbehave, since the costs of playing against the rules of the network will usually be higher than the actual reward of conducting e.g. a "double-spend" attack. The node operators (miners) in, e.g. a PoW network, validate each block and compete with each other. The competition is comparable to a cryptographic puzzle, where all miners compete to be the first to solve the mathematical equation behind the puzzle\cite{berentsen2017}. Only the right unique hash value will solve the puzzle. The first miner that solves the puzzle is allowed to write the transactions on the blockchain thereby creating the next block. In return, the winner earns a \emph{block reward} for the costs incurred in form of the underyling network token. 

This means that all network participants that work towards adding blocks of transactions to the ledger can potentially earn network tokens. By participating in this competition, miners collectively make sure that all transactions included in a block are valid \cite{shermin2019}.

\subsection{Economics to incentivize the evolution towards desired properties} 
\label{sec:economics}


The Psychology of Human Misjudgment, a speech given in 1995 by Charlie Munger \cite{munger2020}, illustrated how behavioral psychology can be applied to business, economics and problem-solving. C. Munger showed how psychology can be used to obtain more structured and thorough understanding of how incentives shape human behavior\cite{munger1995}. 

The basic “\emph{law of behavior}” is that higher incentives will lead to more effort and higher performance \cite{falk2002}. In recent years, also due to the rise of cryptoeconomic networks, the use of incentives in behavioral interaction has become more and more popular. Sometimes the solution to a behavior problem is simply to review and adapt incentives to make sure they align with the desired goal. The effects of incentives depend on how they are designed, the form in which they are given (especially monetary or non-monetary), how they interact with intrinsic motivations and social motivations, and what happens after they are withdrawn. Incentives do matter, but in various and sometimes unexpected ways \cite{gneezy2011}.

Tokenomics encompasses the concept of economic system and optimization design to incentivize specific behaviors in a community, using tokens as the incentive instrument. Tokenomics can now be used to properly design the desired market behavior illustrated in Section \ref{coalition}. This can include predictably failing in certain situations and knowing limitations. This means, the underlying protocol and incentive mechanism design aligns stakeholder interests in the absence of an intermediary in order to create more efficient markets. The challenge, then, is to design the network in such a way that if all participants behave in their own self-interest, the shared goal is achieved, almost as a side-effect\cite{lee2019}.

The more complex and larger the networks are, consisting of different stakeholders, the more important and significant is the design of game theoretical mechanisms. Cryptography can be seen as a linear component, whereas game theory rather an exponential and complex one.

\section{Novel collaboration frameworks}
\label{digitalization}

Conventional cooperation frameworks for cooperative orchestrated networks are mostly hierarchical structured and perform manual processes that can hardly meet the requirements of international collaborations with a multitude of stakeholders\cite{axelrod}. They usually lack the necessary degree of consistently fair co-determination rules and trust, since hierarchical structures are very susceptible to individuals making decisions that are not in the public interest (e.g. moral hazard). DLT-based collaboration frameworks, such as \emph{Decentralized (Autonomous) Organizations} (DAOs)\cite{dao2020}, are now being explored that enhance co-determination possiblities for all stakeholders of an ecosystem. The goal of a DAO is to create an organization that can function without “human” hierarchical management \cite{hassan2021}.

On the other side, if everyone is involved in the decision-making process, efficiency and scalability suffer and the collaboration is literally paralysed as the number of participants increases. Therfore it exists a tension between scalability, i.e. the number of decisions that a collective can make in a given period of time, and resilience, i.e. the incorruptibility of these decisions \cite{ostrom1990}. In addition, the success of open innovation initiatives often fails due to the lack of collective oriented incentive mechanisms to motivate the collective effort required to succeed with such initiatives \cite{breunig2014}.

The goal is to shift from closed, Web2 monopol-centric coordination to federated and distributed protocol-centric coordination.

\subsection{Why Conventional approaches fail}
\label{sec:fail}

Conventional approaches to build a commons-based digital infrastructure have so far failed in the absence of strong incentives. An example would be Quaero \cite{quaero2019}, which was a European research and development program initiated in 2005 aiming to realize a European search engine surpassing American-based ones. Critics immediately realised that “Going head-to-head with Google with a project involving well-funded, energetic entrepreneurs would be foolish. Attempting the same with a multi-government collaboration is beyond description.”\cite{ross2007}. In 2013 Quaero was aborted after spending more than 200M\EUR of public funds \cite{ec2008}\cite{quaero2013}.

\begin{figure*}[h!]
	\centering
	\includegraphics[width=11.2cm, height=5cm]{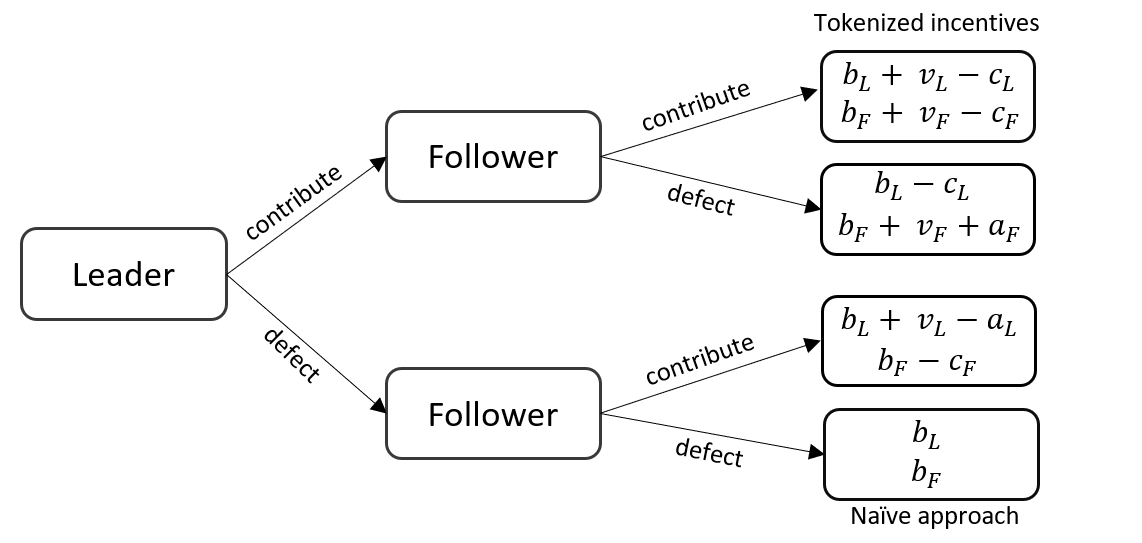}
	\caption{Knowledge Contribution Game to model building process of commons-based infrastructure}
	\label{fig:cgame}
\end{figure*}

The problem of building commons based digital infrastructures can be thought of as a “Knowledge Contribution Game" \cite{gaechter2010}, where two parties called “Leader” and “Follower” iteratively decide whether to contribute to the digital infrastructure or defect/free-ride\footnote{The free-rider problem is a type of market failure that occurs when those who benefit from public resources do not pay for them.}. The generic properties of such conflicts of interest in knowledge sharing can be illustrated in a sequential dynamic relationship. Figure \ref{fig:cgame} illustrates such a knowledge contribution game. 

For illustration purposes, we call the players \emph{Leader} and \emph{Follower}. Both players have the choice of contributing to share their knowledge, \emph{contribute}, or defecting to share knowledge, \emph{defect}. In our example in Figure \ref{fig:cgame}, the leader moves first and decides whether to share or conceal his knowledge. The Follower is subsequently informed of the Leaders choice and decides whether to share or defect. The illustrated model even allows the Follower to decide whether or not to contribute to share knowledge, even if the Leader has decided to defect. After Follower's choice, the game ends and payoffs are realized. \(b_i (i=L,F)\) denotes a basis payoff where \(v_i\) is the value enhancement through sharing knowledge from other player's, \(a_i\) is the exclusivity payoff from appropriating other player's contribution, and \(k\) denotes the expenses for sharing explicit knowledge.

Figure \ref{fig:cgame} illustrates that knowledge sharing is a coordination game with multiple equilibria, since there a different outcomes of knowledge sharing as illustrated in our example. These different outcomes allow us to model the fundamental structures of knowledge sharing (sequentiality and asymmetry in conflicts of interest). In the absence of central coordination, incentive structure and other contextual factors according to social preferences, such as fairness, efficiency seeking, and reciprocity influence the players. Sharing knowledge entails opportunity costs of giving up exclusivity that must be added to the possible costs of sharing knowledge \cite{gaechter2010}. Nevertheless, Song et.\,al (2008) have shown that sharing of mutually-complementary knowledge resources across organizations in a competitive strategic alliance, makes every alliance member's expected income increase which is the reason why this becomse the source of power in establishing an alliance \cite{song2008}. 

Such a knowledge game is very similar to the mechanics of \emph{Open Source Software} (OSS) projects, such as collaborative data spaces (CDS), and provides two fundamental insights \cite{alventosa2018}:

\begin{enumerate}
	\item If leaving uncontrolled, the conventional approach will result in both parties defecting. Such an outcome can be seen in some OSS-projects.
	\item If it turns out to be economically beneficial, then mutual contribution will definitely happen. Such an outcome can be achieved by implementing the right incentives.
\end{enumerate}

In this respect, it is key to realize that a common and shared strategic goal is necessary but not sufficient to avoid the \emph{free-rider problem}. The free-rider problem can temporarily alleviated by public funding but as this funding is based on the work done and not the results achieved, the incentive to provide a long-term operating infrastructure is limited. The rationale behind this is to link funding more directly with outputs and outcomes, rather than inputs and processes. This provides incentives to improve project effectivness and increases accountability for the agents \cite{oecd2014}. 
  
This is especially true, if no long-term subsidizing by public authorities is planned. Long-term incentive alignment play a crucial role to ensure long-term goal orientated cooperation and contribution within a stakeholder group \cite{volden2018}.


\subsection{Digitalization of governance}
\label{governance}

The coordination of the participants therefore requires a well-considered design approach and a collection of sophisticated rules and processes, which are summarized under the term "\emph{governance}" \cite{ostrom1994}. Governance is an applied social problem and generally refers to the process of reaching social consensus. 

Humans have organized and then reorganized themselves under different schemas across time, and these organizational schemas tend to be more and more socially scalable. Guidelines, decision-making and conflict-resolution processes for cooperation can take different forms in its degree of formalization and ways of implementation \cite{chootray2009}. Today's cooperations are build on analog contracts or informal agreements which includes a set of exchange conditions. Those are manually interpreted and enforced by legal action at a higher instance like e.g. a nation state court. Machine-readable and verifiable contracts and digital tokens as a representation of rights and tool for incentives provides the foundation to digitized governance processes across multi un-trusted parties \cite{john2018}. 

An example demonstrate DAOs. A DAO is an implementation of cooperation that lives autonomously in virtual space. It leverages social networks combined with DLT as a collaboration tool. The organizational structure and processes for decision-making are digitally mapped without the need of a central intermediary. Trust is created in software and processes through open source code and targeted contributions are encouraged through token-based incentive mechanisms \cite{liu2020}.

The essential feature of DAOs is that their operating rules are programmed, meaning that they are automatically applied and enforced when the conditions specified in the protocol are met. This differentiates them from traditional organizations, whose rules form guidelines that someone, mostly a human, must interpret and apply. Therefore, DAOs efficiency in regard to decision-making is obvious higher in comparison to traditional governed organizations \cite{honigman2019}. 

Within a DAO, the decision-making process can be reviewed and audited at any time. In this way, it provides a feasible solution to audit the process and significantly improves the trustworthiness in comparison to centralized alternatives. Cryptographic token are involved to orchestrate e.g. the voting process and voting weights of its participants.

The concept of a DAO fails if it becomes centralized. Yet centralization in governance is perhaps the largest threat to a DAO. With increasing centralization comes increasing risk of an organization's unifying principles changing or being ignored over time. Decentralization forces the shared understanding and shared social contract to be of primary importance for the life of the organization \cite{kaal2021}. DAOs currently still raise legal and technical questions, but they provide an outlook on how contracts and cooperation processes may be gradually transferred to the digital age \cite{buech2019}. 

Such collaboration frameworks leverage DLT to offer an alternative for more fluid organizational structures and can be described as novel approach for borderless cooperation in the digital age and are well suited for CDS \cite{buterin2014}.

\section{From incentives to tokenized incentives} 

Understanding incentives is key to understand people and socio-economy. Vice versa, failing to recognize the importance of incentives can lead to misconstructions and errors \cite{falk2002}. Incentive mechanisms can be described as rules of the game for groups of individuals which are designed in such a way, that certain goals are achieved if the group members act rationally on their own benefit within the framework of these rules. The desired social state is then compatible with the individual incentives \cite{buterin2017}. The group can be a society, an organisation or a community of contractual partners such as CDS initiatives like \emph{GAIA-X}.

Incentives can be either:
\begin{enumerate}
	\item Negative incentives (\emph{control}): Designed to increase willingness to show the desired behavior by sanctioning its undesired counterpart.
	\item Positive incentives (\emph{enabler}): Designed to enable and increase willingness to show the desired behavior by rewarding it\cite{benabou2006}.
\end{enumerate}

Incentive mechanisms are a conceptual tool of economic and game-theory. They are used for economic analysis in various fields of economics, e.g. incentive mechanisms in the taxation of industrial economics, the allocation of goods, resources and risks or the economics of the public sector. Incentives can be any kind of reward or punishment that range from remunerative to moral or reputational aspects \cite{samson2014}.

Incentive mechanisms serve to implement the given social goal by a non-cooperative balance of the resulting game. The term “non-cooperative” means this branch of game theory explicitly models the process of players making choices out of their own interest. Cooperation can, and often does, arise in non-cooperative models of games, when players find it in their own best interests \cite{chandrasekaran1994}.

\subsection{Monetary incentives}

Money is one of the greatest forces of social good, besides social norms. Remunerative incentives should be integrated into socio-economic interaction to achieve the desired behaviour. Punishment, e.g. through taxes, works good to prevent actions whereas incentives work best to encourage them \cite{tabellini2008}. 

For simplicity, we solely focus on monetary incentives in this primer. Moral and coercive incentives are important, but less convenient to illustrate. Remunerative incentives, analogue to monetary-incentives, lay at the heart of tokenized networks. Those incentives provide the operators of the network with financial rewards in order to sustain the network, e.g. in the Bitcoin blockchain the so-called block rewards \cite{nakamoto2008}.

The most common financial incentive form of tokens are \emph{Initial Coin Offerings} (ICO) and are known in the Web3 community. ICO's are used to fund and bootstrap the development of different platform cooperatives or protocol applications (e.g. dApps). ICO's can be compared to traditional \emph{Initial Public Offerings} (IPO), but offers additional advantages e.g. direct democratic participation. During ICOs, organizations distribute their tokens to investors in exchange for capital. Investors become token holders and provide different functions and utilities within the issuer's network as soon as the project is launched \cite{schuckes2020}. Tokens incentivize to deliver a viable product that accrues value and holds in the future. 

The funding amounts in ICOs exceed most investment rounds by traditional funding vehicles. The advantage of this type of financing over conventional contribution-financed joint ventures is the directly monetizable benefit, which is reflected in the value of the token. Recent data shows an average of 1600 investors per ICO, with average funds raised of \$9 million. Conventional crowdfunding is dramatically smaller and much less internationalised \cite{boreiko2020}.

In addition to having a novel financing mechanism, tokens help to address the coordination problem that is common in network adoption. With tokens, the platform trades off future revenue for present revenue, which helps solving the coordination problem \cite{bakos2018}. 

\subsection{Tokenized incentives}

From mining rewards, to transaction fee-settings, to prediction markets, \emph{tokenized} incentives are omnipresent in DLT-based protocols. Cryptographic tokens are digital incentive mechanisms and a central component of the solution to the coordination problem. By means of tokens, one can create efficient alignment of different stakeholders within a digital network \cite{bakos2018}. By this, ecosystems can incentivize the right behavior for each stakeholder in order to achieve the desired system-level behavior. \emph{Tokenized incentives} are used to orchestrate the creation and governing of the evolution of such protocols to build incentive-aligned and robust P2P networks \cite{buterin2014}. 

Token design is a complex and time-consuming endeavor that includes social choice, financial and legal aspects \cite{buech2019}. Tokens, properly designed, represent the ownership of scarce digital resources and coordinate the actors in a given network, e.g. through block rewards in the Bitcoin network \cite{Mersch2020}. DLT-based protocols provide a new way of issuing, redeeming and automatically enforcing the rights associated to these tokens, in a digital and distributed manner. It is therefore economically irrational for a participant to disregard the established rules, as the economic benefit to behave against the rules leads to a lower surplus for the individual \cite{garay2015}.

To sum up, incentive-aligned systems linked to an underlying cryptographic token, enable the coordination and allocation of resources from the start of a project towards the desired properties \cite{falk2002}, more than this, incentive-aligned networks even allow higher socio-economic efficiency, as illustrated in Section \ref{coalition}.

\subsection{Tokenize to make it a success}
\label{sec:success}

If we combine trustless technologies, such as \emph{DLT}, the \emph{coopetition} approach mentioned in Section \ref{coalition}, novel collaboration frameworks mentioned in Section \ref{governance} and digital incentives in the form of \emph{token} in Section \ref{tokenomics}, we have all necessary tools to design and create robust, incentive-aligned collaboration networks. Those network offer an alternative for more fluid organizational structures and can be described as novel approach for borderless cooperation in the digital age \cite{buterin2014}. 


Incentive mechanism design is a critical part of the overall economic design of such networks and also a crucial element when it comes to the success of public funded initiatives. It is the piece that enables a platform's value proposition and structures the system for which the token of the platform will be designed. From an economics perspective, it is the crux of the system.

Incentive-efficient funding increases the so-called leverage effect of funding instruments, e.g. the effect of subsidies on private expenditure as an input for research and development activities \cite{oecd2020}. The reason for this is the simultaneous divergence of interests and information between potential recipients and funding agencies. In principle, funding policy can therefore be approached at two levels, namely through:

\begin{enumerate}
	\item Reducing the information asymmetry, i.e. the alignment of the divergent information situations of the funding recipient and the funding agency \emph{or}
	\item Incentive mechanisms that align the interests of funding recipients with those of the (welfare-maximizing) funding provider
\end{enumerate}

By interlinking the two interest groups, principal and agent, the behaviour of the agent can be controlled. This agent is to be encouraged to efficiently perform the contractually agreed and owed service and not to deceive the principal either before or after contract conclusion \cite{Klement2004}. A well-balanced token design would incentivize network participants to take a risk in adopting a new platform before it is clear that it is worth it, and reward them with ownership that will have future value thanks to their contribution. Interlinking principal and agent on basis of incentive alignment through the use of cryptographic tokens, allows a target-oriented and efficient funding project progress and will foster the success of CDS in the digital age.

    \section{Conclusion}

Collaborative Data Spaces based on DLT and fueled by digitized incentives in the form of cryptographic token can be regarded as the next evolutionary step of digital collaboration. DLT provides protocols for novel collaboration frameworks that provide higher levels of transparency and efficiency while reducing bureaucracy with self-enforcing code \cite{buterin2014}. 

Thus, cryptographic tokens are far more than financial speculation instruments. We have discussed how a token can represent a multitude of aspects, such as ownership or rights of participation, and are an integral part of the underlying network. Such tokens can be leveraged for the coordination, optimization and governance of large networks at scale in a decentralized manner. These properties make them an indispensable tool to build fair and efficient digital ecosystems.

Tokenomics is posed to speed up the evolution of DLT-based ecosystems where tokenized incentives align multiple interests alongside a desired outcome, from the funding towards the desired system properties. Tokenomics therefore assures a long-term perspective to \emph{coopetitive} collaboration in the digital age. 

Adherence to the principles and values that are designed and set through the evolutionary process of a digital collaboration can lead to a functioning market economy characterized by socio-economic efficiency. Efficiency by means of resource allocation provides better socio-economic output in the sense for each individual and for the digital cooperative as a whole. Figuratively, for an emerging digital economy, this means that the use of DLT and tokenized incentives can lead to a market economy characterized by socio-economic efficiency.





\bibliographystyle{plain}
\bibliography{01_content/00_references}

\begin{thebibliography}{10}

\bibitem{akerlof1970}
G.~A. Akerlof.
\newblock The market for "lemons": Quality uncertainty and the market
  mechanism.
\newblock pages 448--500, 1970.

\bibitem{altman2006}
D.~Altman.
\newblock Q \& answers with joseph e. stiglitz.
\newblock 2006.

\bibitem{alventosa2018}
A.~Alventosa and P.~Hernández.
\newblock {\em Coordination Concerns: Concealing the Free Rider Problem}.
\newblock 2018.
\newblock Available at
  \url{https://www.intechopen.com/books/game-theory-applications-in-logistics-and-economy/coordination-concerns-concealing-the-free-rider-problem}.

\bibitem{atzori2017}
M.~Atzori.
\newblock Blockchain technology and decentralized governance: Is the state
  still necessary?
\newblock 2017.
\newblock Available at \url{http://dx.doi.org/10.22495/jgr_v6_i1_p5}.

\bibitem{axelrod}
R.~Axelrod.
\newblock The complexity of cooperation: Agent-based models of competition and
  collaboration.
\newblock 1997.

\bibitem{bakos2018}
Y.~Bakos and H.~Halaburda.
\newblock The role of cryptographic tokens and icos in fostering platform
  adoption.
\newblock 2018.

\bibitem{benos2017}
E.~Benos, R.~Garratt, and P.~Gurrol-Perez.
\newblock The economics of distributed ledger technology for securities
  settlement.
\newblock 2017.

\bibitem{berentsen2017}
A.~Berentsen and F.~Schär.
\newblock {\em Bitcoin, Blockchain und Kryptoassets: Eine umfassende
  Einführung}.
\newblock Books on Demand, 2017.

\bibitem{blankart2000}
C.~B. Blankart.
\newblock The process of government centralization: A constitutional view.
\newblock 2000.
\newblock Available at
  \url{https://link.springer.com/article/10.1023%2FA%3A1009018032437}.

\bibitem{ids2018}
Menz~N. Bohlen~V., Bruns~L.
\newblock Open data spaces. towards the ids open data ecosystem., 2020.
\newblock Available at \url{https://doi.org/10.5281zenodo.5675977}.

\bibitem{boreiko2020}
D.~Boreiko and D.~Risteski.
\newblock Serial and large investors in initial coin offerings.
\newblock 2020.

\bibitem{breunig2014}
K.~Breunig, T.~Aas, and K.~Hydle.
\newblock Incentives and performance measures for open innovation practices.
\newblock 18, 2014.

\bibitem{burniske2019}
C.~Burniske.
\newblock Protocols as minimally extractive coordinators.
\newblock 2019.
\newblock Available at
  \url{https://www.placeholder.vc/blog/2019/10/6/protocols-as-minimally-extractive-coordinators}.

\bibitem{buterin2014}
V.~Buterin.
\newblock Daos, dacs, das and more: An incomplete terminology guide.
\newblock 2014.
\newblock Available at
  \url{https://blog.ethereum.org/2014/05/06/daos-dacs-das-and-more-an-incomplete-terminology-guide/}.

\bibitem{buterin2017}
V.~Buterin.
\newblock Introduction to cryptoeconomics.
\newblock 2017.
\newblock Available at
  \url{https://crypto.berlin/introduction-to-cryptoeconomics-by-vitalik-buterin}.

\bibitem{benabou2006}
R.~Bénabou and J.~Tirole.
\newblock Incentives and prosocial behavior.
\newblock pages 1652--1678, 2006.

\bibitem{buech2019}
M.~Büch.
\newblock Die idee der dao - von missverständnissen und potentialen.
\newblock 2019.
\newblock Available at
  \url{https://www.btc-echo.de/die-idee-der-dao-von-missverstaendnissen-und-potentialen/}.

\bibitem{chandrasekaran1994}
R.~Chandrasekaran.
\newblock Cooperative game theory.
\newblock 1994.
\newblock Available at
  \url{www.utdallas.edu/chandra/documents/6311/coopgames.pdf}.

\bibitem{chesbrough2021}
H.~Chesbrough and A.~Radziwon.
\newblock International data spaces: A collaborative organizational moonshot.
\newblock 2021.

\bibitem{chootray2009}
V.~Chootray and G.~Stoker.
\newblock Governance theory and practice: A cross-disciplinary approach.
\newblock 2009.

\bibitem{christakis2021}
T.~Christakis.
\newblock European digital sovereignty: Successfully navigating between the
  "brussels effect" and europe's quest for strategic autonomy.
\newblock 2020.

\bibitem{protocol2020}
P.~Christensson.
\newblock Protocol definition.
\newblock 2020.
\newblock Available at \url{https://techterms.com/definition/protocol}.

\bibitem{ec2008}
European Commission.
\newblock State aid: Commission authorises aid of 99 million eur to france for
  quaero programme.
\newblock 2008.
\newblock Available at
  \url{https://ec.europa.eu/commission/presscorner/detail/en/IP_08_418}.

\bibitem{crosby2016}
M.~Crosby, P.~Pattanayak, S.~Verma, and V.~Kalyanaraman.
\newblock Blockchain technology: Beyond bitcoin. applied innovation.
\newblock 2016.

\bibitem{falk2002}
E.~Fehr and A.~Falk.
\newblock Psychological foundations of incentives.
\newblock pages 687--693, 2002.

\bibitem{floros2019}
E.~J. Floros.
\newblock Web 3.0 - the internet of value.
\newblock 2019.
\newblock Available at \url{https://doi.org/10.1002/9781119551973.ch38}.

\bibitem{fox2019}
E.~M. Fox.
\newblock Platforms, power and the antitrust challenge: A modest proposal to
  narrow the u.s. - europe divide.
\newblock 2019.
\newblock Available at \url{https://digitalcommons.unl.edu/nlr/vol98/iss2/4}.

\bibitem{gaia2022}
{GAIA-X}.
\newblock Data infrastructure, 2022.
\newblock Available at
  \url{https://www.data-infrastructure.eu/GAIAX/Navigation/EN/Home/home.html}.

\bibitem{galloway2021}
S.~Galloway.
\newblock Scarcity cred.
\newblock 2021.
\newblock Available at \url{https://www.profgalloway.com/scarcity-cred/}.

\bibitem{garay2015}
J.~A. Garay, A.~Kiayias, and N.~Leonardos.
\newblock The bitcoin backbone protocol: Analysis and applications.
\newblock 2015.

\bibitem{gehl2012}
R.~Gehl.
\newblock Distributed centralization: Web 2.0 as a portal into users’ lives.
\newblock 2012.
\newblock Available at
  \url{https://csalateral.org/issue/1/distributed-centralization-web-2-0-portal-users-gehl/}.

\bibitem{gneezy2011}
U.~Gneezy, S.~Meier, and P.~Rey-Biel.
\newblock When and why incentives (don’t) work to modify behavior.
\newblock pages 192--210, 2011.

\bibitem{granot2019}
E.~Granot.
\newblock On the origin of the value of cryptocurrencies.
\newblock 2019.
\newblock Available at \url{https://doi.org/10.5772/intechopen.79451}.

\bibitem{gaechter2010}
S.~Gächter, G.~von Krogh, and S.~Haefliger.
\newblock Initiating private-collective innovation: The fragility of knowledge
  sharing.
\newblock 2010.
\newblock Available at
  \url{https://linkinghub.elsevier.com/retrieve/pii/S0048733310001149}.

\bibitem{han2012}
Z.~Han, D.~Niyato, W.~Saad, T.~Basar, and A.~Hjørungnes.
\newblock Game theory in wireless and communication networks: theory, models,
  and applications.
\newblock 2012.
\newblock Available at
  \url{https://www.researchgate.net/publication/267132783_Game_Theory_in_Wireless_and_Communication_Networks_Theory_Models_and_Applications}.

\bibitem{Harz2018}
D.~Harz and M.~Boman.
\newblock The scalability of trustless trust.
\newblock 2018.
\newblock Available at \url{https://arxiv.org/abs/1801.09535}.

\bibitem{hassan2021}
S.~Hassan and P.~De Filippi.
\newblock Decentralized autonomous organization.
\newblock 2021.

\bibitem{honigman2019}
P.~Honigman.
\newblock The spring of the dao's.
\newblock 2019.
\newblock Available at \url{https://hackernoon.com/what-is-a-dao-c7e84aa1bd69}.

\bibitem{horne2018}
J.~Horne.
\newblock The emergence of cryptoeconomic primitives.
\newblock 2018.
\newblock Available at
  \url{https://blog.coinbase.com/the-emergence-of-cryptoeconomic-primitives-14ef3300cc10}.

\bibitem{ibm2015}
IBM.
\newblock The economy of things: Extracting new value from the internet of
  things.
\newblock 2015.

\bibitem{john2018}
T.~John and M.~Pam.
\newblock Complex adaptive blockchain governance.
\newblock 2018.
\newblock Available at \url{https://doi.org/10.1051/matecconf/201822301010}.

\bibitem{kaal2021}
W.~A. Kaal.
\newblock A decentralized autonomous organization (dao) of daos.
\newblock 2021.
\newblock Available at
  \url{https://papers.ssrn.com/sol3/papers.cfm?abstract_id=3799320}.

\bibitem{khan2002}
A.~Khan and Y.~Sun.
\newblock Non-cooperative games with many players: Handbook of game theory with
  economic applications.
\newblock pages 1761--1808, 2002.
\newblock Available at
  \url{https://www.elsevier.com/books/handbook-of-game-theory-with-economic-applications/aumann/978-0-444-89428-1}.

\bibitem{Klement2004}
B.~Klement.
\newblock Ökonomische kriterien und anreizmechanismen für eine effiziente
  förderung von industrieller forschung und innovation mit einer empirischen
  quantifizierung der hebeleffekte von förderinstrumenten in Österreich.
\newblock 2004.
\newblock Available at \url{https://core.ac.uk/reader/11007582}.

\bibitem{kroer2014}
C.~Kroer and T.~Sandholm.
\newblock Extensive-form game abstraction with bounds.
\newblock EC '14, page 621–638. Association for Computing Machinery, 2014.

\bibitem{kuznetsov2017}
N.~Kuznetsov.
\newblock Decentralizing the world - blockchain and the removal of centralized
  profit centers.
\newblock 2017.
\newblock Available at
  \url{https://www.forbes.com/sites/nikolaikuznetsov/2017/08/16/decentralizing-the-world-blockchain-and-the-removal-of-centralized-profit-centers/?sh=4e80ccd753f4}.

\bibitem{lee2019}
J.~Y. Lee.
\newblock A decentralized token economy: How blockchain and cryptocurrency can
  revolutionize business.
\newblock 2019.

\bibitem{liu2020}
L.~Liu, S.~Zhou, H.~Huang, and Z.~Zheng.
\newblock From technology to society: An overview of blockchain-based dao.
\newblock 2010.
\newblock Available at \url{https://arxiv.org/abs/2011.14940}.

\bibitem{marhsall1976}
J.~M. Marshall.
\newblock Moral hazard.
\newblock 66:880--890, 1976.

\bibitem{Mersch2020}
C.~Mersch.
\newblock An (entrepreneurial) investor’s take on the utility of tokens
  beyond payment.
\newblock 2020.

\bibitem{micro2019}
Microsoft.
\newblock Tokenization: Establishing digital representations of value as the
  medium of exchange.
\newblock 2019.

\bibitem{minarsch2020}
D.~Minarsch, M.~Favorito S.~A.~Hosseini, and J.~Ward.
\newblock Autonomous economic agents as a second layer technology for
  blockchains: Framework introduction and use-case demonstration.
\newblock 2020.

\bibitem{mohanta2018}
B.~K. Mohanta, S.~S. Panda, and D.~Jena.
\newblock An overview of smart contract and use cases in blockchain technology.
\newblock pages 1--4, 2018.

\bibitem{munger1995}
C.~Munger.
\newblock The psychology of human misjudgment, 1995.
\newblock Available at \url{https://www.youtube.com/watch?v=pqzcCfUglws}.

\bibitem{nakamoto2008}
S.~Nakamoto.
\newblock Bitcoin: A peer-to-peer electronic cash system.
\newblock 2009.
\newblock Available at \url{http://bitcoin.org/bitcoin.pdf}.

\bibitem{nalebuff2021}
B.~Nalebuff and A.~Brandenburger.
\newblock Co-opetition: competitive and cooperative business strategies for the
  digital economy.
\newblock 1997.

\bibitem{oecd2014}
OECD.
\newblock Technical workshop on results-based funding, workshop report.
\newblock 2014.
\newblock Available at
  \url{https://www.oecd.org/dac/results-development/technicalworkshoponresultsbasedfunding.htm}.

\bibitem{oliveira2018}
L.~Oliveira, L.~Zavolokina, and I.~Bauer.
\newblock To token or not to token: Tools for understanding blockchain tokens.
\newblock 2018.

\bibitem{ostrom1990}
E.~Ostrom.
\newblock {\em Governing the Commons: The Evolution of Institutions for
  Collective Action}.
\newblock Cambridge University Press, 1990.

\bibitem{ostrom1994}
E.~Ostrom, R.~Gardner, and J.~Walker.
\newblock {\em Rules, Games, and Common-Pool Resources}.
\newblock University of Michigan Press, 1994.
\newblock Available at \url{https://www.press.umich.edu//9745}.

\bibitem{Otto2019}
B.~Otto, D.~Lis, J.~Jürjens, S.~Opriel, F.~Howar, and S.~Meister.
\newblock Data ecosystems: Conceptual foundations, constituents and
  recommendations for action.
\newblock 2019.

\bibitem{poddey2019}
A.~Poddey and N.~Scharmann.
\newblock On the importance of system-view centric validation for the design
  and operation of a crypto-based digital economy.
\newblock 2019.
\newblock Available at \url{https://arxiv.org/abs/1908.08675}.

\bibitem{polanyi2001}
K.~Polanyi.
\newblock The great transformation: The political and economic origins of our
  time author.
\newblock 2001.

\bibitem{quaero2013}
Quaero.
\newblock Le programme quaero s'acheve.
\newblock 2013.
\newblock Available at
  \url{http://www.quaero.org/31-decembre-2013-le-programme-quaero-sacheve/}.

\bibitem{ross2007}
P.~Ross.
\newblock What's the latin for "delusional"?
\newblock 2007.
\newblock Available at \url{https://ieeexplore.ieee.org/document/4049963}.

\bibitem{samieh2007}
H.~Samieh and K.~Wahba.
\newblock Knowledge sharing behavior from game theory and socio-psychology
  perspectives.
\newblock 2007.

\bibitem{samson2014}
A.~Samson.
\newblock {\em The Behavioral Economics Guide - 2014}.
\newblock 2014.

\bibitem{santeri2018}
P.~Santeri and N.~Pekka.
\newblock Security and privacy challenges and potential solutions for dlt based
  iot systems.
\newblock In {\em 2018 Global Internet of Things Summit (GIoTS)}, pages 1--6,
  2018.

\bibitem{schuckes2020}
M.~Schückes and T.~Gutmann.
\newblock Why do startups pursue initial coin offerings? the role of economic
  drivers and social identity on funding choice.
\newblock 2020.
\newblock Available at \url{https://doi.org/10.1007/s11187-020-00337-9}.

\bibitem{oecd2020}
OECD Blockchain~Policy Series.
\newblock The tokenisation of assets and potential implications for financial
  markets.
\newblock 2020.
\newblock Available at
  \url{{www.oecd.org/finance/The-Tokenisation-of-Assets-and-Potential-
  Implications-for-Financial-Markets.htm}}.

\bibitem{shapley1953}
L.~S. Shapley.
\newblock A value for n-person games.
\newblock 1953.

\bibitem{pos2017}
J.~Siim.
\newblock Proof-of-stake, 2017.
\newblock In: Research seminar in cryptography.

\bibitem{smith1759}
A.~Smith.
\newblock The theory of moral sentiments.
\newblock 1759.

\bibitem{song2008}
X.~Song, J.~Li, and N.~Xu.
\newblock Dynamic game analysis on knowledge-sharing and knowledge-spillover in
  competitive alliances.
\newblock pages 782--788, 2008.

\bibitem{stiglitz1989}
J.~E. Stiglitz.
\newblock Principal and agent.
\newblock 1989.

\bibitem{stiglitz2002}
J.~E. Stiglitz.
\newblock There is no invisible hand.
\newblock 2002.

\bibitem{tabellini2008}
G.~Tabellini.
\newblock The scope of cooperation: Values and incentives.
\newblock pages 905--950, 2009.
\newblock Available at \url{https://doi.org/10.1162/qjec.2008.123.3.905}.

\bibitem{tasca2020}
P.~Tasca.
\newblock Internet of value: A risky necessity.
\newblock 2020.
\newblock Available at
  \url{https://www.frontiersin.org/article/10.3389/fbloc.2020.00039 }.

\bibitem{tinsman2018}
B.~Tinsman.
\newblock What is tokenomics?
\newblock 2018.
\newblock Available at
  \url{https://www.he3labs.com/blog/2018/6/4/what-is-tokenomics}.

\bibitem{volden2018}
G.~H. Volden.
\newblock Public funding, perverse incentives and counterproductive outcomes.
\newblock 2018.
\newblock Available at
  \url{https://www.emerald.com/insight/content/doi/10.1108/IJMPB-12-2017-0164/full/html}.

\bibitem{shermin2019}
S.~Voshmgir.
\newblock {\em Token Economy}.
\newblock Blockchain Hub Berlin, 2019.

\bibitem{zargham2020}
S.~Voshmgir and M.~Zargham.
\newblock Foundations of cryptoeconomic systems.
\newblock 2020.
\newblock Available at \url{https://epub.wu.ac.at/7309/}.

\bibitem{quaero2019}
{Wikipedia}.
\newblock Quaero, 2019.
\newblock Available at \url{https://en.wikipedia.org/wiki/Quaero}.

\bibitem{munger2020}
{Wikipedia}.
\newblock C. munger, 2020.
\newblock Available at \url{https://en.wikipedia.org/wiki/Charlie\_Munger}.

\bibitem{dao2020}
{Wikipedia}.
\newblock Decentralized autonomous organization, 2020.
\newblock Available at
  \url{https://en.wikipedia.org/wiki/Decentralized_autonomous_organization}.

\bibitem{wright2015}
A.~Wright and P.~De Filippi.
\newblock Decentralized blockchain technology and the rise of lex
  cryptographia.
\newblock 2015.
\newblock Available at \url{https://ssrn.com/abstract=2580664}.

\bibitem{Survey2019}
L.~Ziyao, C.~Ngyuen, W.~Wang, N.~Dusit, L.~Ying-Chang W.~Ping, and K.~Dong.
\newblock A survey on applications of game theory in blockchain.
\newblock 2019.
\newblock Available at \url{https://arxiv.org/abs/1902.10865}.

\bibitem{zumbansen2007}
P.~Zumbansen.
\newblock The law of society: Governance through contract.
\newblock 2007.

\end{thebibliography}




\end{document}